\documentclass[11pt,reqno]{article}
\newif\ifclean
\cleantrue  
\usepackage{graphicx}
\usepackage{amsmath}
\usepackage{amssymb}
\usepackage{subfig}
\usepackage{geometry} 
\graphicspath{{{./}}}

\usepackage{color}
\newcommand{\comment}[1]{\textcolor{cyan}{{[ \sc{#1} ]}}} 
\newcommand{\red}[1]{\textcolor{red}{{#1}}}
\newcommand{\caution}{\red{\bf Draft: \today. Do not distribute.}}
\ifclean
\renewcommand{\comment}[1]{{}}
\else
\fi

\newcommand{\fref}[1]{Fig.\,\ref{#1}}

\newcommand{\sref}[1]{Sec.\!~\ref{#1}}

\newcommand{\cref}[1]{Ref.\,\cite{#1}}
\newcommand{\crefs}[1]{Refs.\,\cite{#1}}

\newcommand{\ie}{{\it i.e.}\!\, }

\newcommand{\etc}{{\it etc.}\! }
\newcommand{\etal}{{\it et al.} }

\newcommand{\Cbb}{\mathbb{C}}

\newcommand{\xb}{\mathbf{x}}
\newcommand{\nb}{\mathbf{n}}
\renewcommand{\sb}{\mathbf{s}}
\newcommand{\Eb}{\mathbf{E}}

\newcommand{\Sb}{\mathbf{S}}

\newcommand{\strain}{{\epsilon}}
\newcommand{\stress}{{\sigma}}

\newcommand{\pth}{\phi}

\title{\bf Prediction of the evolution of the stress field of polycrystals undergoing elastic-plastic deformation with a hybrid neural network model}
\author{Ari Frankel, Kousuke Tachida, Reese Jones\footnote{Corresponding author: \tt rjones@sandia.gov}\\
{\it Sandia National Laboratories, Livermore, CA 94551}
}

\begin{document}
\ifclean
\date{}
\else
\date{\caution}
\fi

\maketitle{}

\begin{abstract}
Crystal plasticity theory is often employed to predict the mesoscopic states of polycrystalline metals, and is well-known to be costly to simulate.
Using a neural network with convolutional layers encoding correlations in time and space, we were able to predict the evolution of the stress field given only the initial microstructure and external loading.
In comparison to our recent work we were able to predict not only the spatial average of the stress response but the field itself.
We show that the stress fields and their rates are in high fidelity with the crystal plasticity data and have no visible artifacts.
Furthermore the distribution stress throughout the elastic to fully plastic transition match the truth provided by held out crystal plasticity data.
Lastly we demonstrate the efficacy of the trained model in material characterization and optimization tasks.
\end{abstract}

\section{Introduction}

The modeling of the mechanical behavior of polycrystalline metals has wide spread technological relevance.
The prediction of the stress states of polycrystals as they deform is difficult especially during transition to plasticity.
Homogenization theory can provide  bounds and approximations to the macroscopic response, but is, by construction, insensitive to the details of the microscopic states influencing the response of a particular polycrystal.
On the other hand, crystal plasticity theory \cite{taylor1934mechanism,kroner1961plastic,bishop1951xlvi,bishop1951cxxviii,mandel1965generalisation,dawson2000computational,roters2010overview} provides a constitutive model that enables detailed meso-scale simulations of metals; however, it is complex and costly to simulate.
Many higher level models of the phenomenology of plastic deformation have been developed over decades \cite{hill1998mathematical,lubliner2008plasticity} and are often calibrated to represent polycrystalline metals.
While this approach is sufficient in the small grain/large sample limit (representative volumes \cite{hill1963elastic,drugan1996micromechanics} and larger), there is a need for accurate, efficient models of the response of aggregates with smaller sample sizes (stochastic volumes \cite{ostoja2006material}).
These models can be used for stand-ins for material design/optimization, and also for uncertainty quantification and other statistical analysis.

Traditional modeling has progressed slowly over decades.
Recently, across a range of commercial and scientific fields, data-driven modeling and, in particular, machine learning \cite{hastie2005elements,goodfellow2016deep} has risen as a tool to form representative models of complex underlying physics without resorting to gross approximations.
Provided a large enough dataset and a trainable model of sufficient and appropriate complexity, a representation of the underlying physics may be learned with high fidelity.

Although nascent, the field of applying machine learning to materials science has garnered sufficient interest to elicit topical reviews, such as \crefs{liu2017materials,dimiduk2018perspectives,bostanabad2018computational}.
There are also a number of notable individual contributions employing a variety of approaches to the tasks of classification and reconstruction/synthesizing microstructures, and predicting their physical response.
Key ingredients in these tasks are selection of the inputs by feature extraction from image, either manually or by the network, and the output, such as classes of the expected microstructures or their stress response.
Convolutional neural networks (CNNs) \cite{krizhevsky2012imagenet} typically play a role in these tasks since they were developed to efficiently process image data.

Image-based transfer learning is an approach that has leveraged developments in computer science, such as Gatys \etal \cite{gatys2015texture}.
In transfer learning  models trained in similar contexts are reused with minimal additional training.
Lubbers \etal \cite{lubbers2017inferring} adapted a convolutional neural network trained for image recognition to form low-dimensional representations of microstructures in order to generate statistical reconstructions.
Li \etal \cite{li2018transfer} also used transfer learning for microstructure reconstruction for a wide range of classes of microstructure.
Clustering is another of the variety of methods employed for classification.
Papanikolaou \etal \cite{papanikolaou2017learning} employed dislocation dynamics data and a clustering algorithm to classify materials based on prior plastic deformation.
Liu \etal \cite{liu2018microstructural} employed a database of microstructures and a clustering algorithm to predict plastic localization.
Using a variety of techniques, such as visual bag of words, texture and shape statistics, and pre-trained convolutional neural networks, support vector machines, clustering, and random forests, Chowdhury \etal \cite{chowdhury2016image} performed feature extraction and classification of dendritic morphologies.

The task of microstructure reconstruction/generation is particularly germane to a variety of materials studies.
Chen and coworkers \cite{xu2015machine,bessa2017framework,bostanabad2016stochastic,bostanabad2016characterization} have produced a considerable body of work in this emerging field, primarily focussed on microstructure reconstruction \cite{bostanabad2016stochastic}.
Xu \etal \cite{xu2015machine} used pre-selected and winnowed image features in a supervised learning process to represent microstructures.
Bostanabad \etal \cite{bostanabad2016characterization,bostanabad2016stochastic} used a Gibbs sampler and a classification tree to construct binary microstructures.
Generative models such as variational autoencoders (VAEs) \cite{kingma2013auto} and generative adversarial networks (GANs) \cite{goodfellow2014generative} have been employed for microstructure reconstruction and optimization.
Cang \etal \cite{cang2018improving} used a VAE to represent granular microstructures and compared its performance to the more traditional approach of using a Markov random field model.
Yang \etal \cite{yang2018microstructural} trained a GAN to provide latent variables that may be used to optimize the properties of a material structure.
Kalidindi and coworkers \cite{deshpande2013application,yabansu2017extraction,cecen2018material,yang2018deep} have also generated numerous papers in the field of microstructure analysis and reconstruction.
Particularly relevant is the work of Cecen \etal \cite{cecen2018material} and Yang \etal \cite{yang2018deep}, which compared more traditional models, based on statistical and physical descriptors, with data-driven methods like principal component analysis (PCA) and compared to a model where a CNN was used to encode the spatial correlations in the elastic response of a two phase material.
They showed showed that the CNN model was superior to the reduced order models using the physical descriptors and regression, as well as classical models based on homogenization theory.

Other work has also focused on predicting the physical response of microstructures.
Niezgoda \cite{dimiduk2018perspectives,yuan2018machine,niezgoda2013novel}  has been active in the field of applying machine learning to materials science, in part with Kalindi \cite{niezgoda2013novel} and also with commercial groups \cite{dimiduk2018perspectives,yuan2018machine}.
Related to the present effort, Yuan \etal \cite{yuan2018machine} used a random forest model with PCA of texture and the underlying model parameters as inputs to predict the history for multiple loading.
Bessa \etal \cite{bessa2017framework} developed a methodology for applying machine learning to modelling physical response with particular focus on sampling of the training data.
Jones \etal \cite{jones2018machine} designed neural network models of plastic plot based on the traditional stress and flow rule framework and key aspects of classical representation theory.
Recently, Frankel \etal \cite{frankel2019predicting} demonstrated the use of convolutional neural networks to predict the elastic modulus of polycrystalline aggregates and combined the CNN architecture with a recurrent neural network to predict the system average stress evolution through yield and plastic flow.
Although the results were, in part, a proof-of-concept of the use of data-driven modeling for crystal plasticity, the architecture was limited to predicting the homogenized stress states and had reduced accuracy as time progressed.

In this work we will demonstrate the use of variants of the hybrid architecture called a convolutional-long short term memory (ConvLSTM \cite{xingjian2015convolutional}) neural network to predict the evolving stress field of deforming polycrystals.
One of the novelties of the design is using images of one type (microstructure) to predict the evolution of images of another type (stress field).
In \sref{sec:data}, we describe the polycrystal dataset and then, in \sref{sec:nn}, we discuss the particular ConvLSTM-based model in the wider context of neural networks and compare variants of the proposed model.
In \sref{sec:results}, we show results for the prediction of the stress state evolution, including error analysis, statistical comparisons, and a material exploration demonstration.
Lastly, we discuss extensions and additional applications in \sref{sec:conclusion}.

\section{Crystal plasticity data} \label{sec:data}

Crystal plasticity (CP) simulations typically consist of a finite element representation of the grain structure, including a mesh defining the grain boundaries and material models informed by the crystal orientation of each grain and the allowed slip systems.
To provide enough training data for our neural network (described in \sref{sec:training}), we employed $>$ 10$^4$ two dimensional (2D) simulations of polycrystals.
Each consisted of the tensile stress response of 1 $\mu$m$^2$ squares of polycrystalline steel with a single face-centered cubic (FCC) phase.
Each grain was assigned a viscoplastic constitutive model with power-law hardening and the crystal interactions in aggregate were governed by compatibility.

\subsection{Realizations}
To produce the 2D realizations, we utilized the well-tested algorithms in the DREAM.3D software package \cite{Groeber2014} for creating 3D polycrystals.
In particular, we created 1 $\mu$m$^2$ cubes with a realistic polycrystalline morphology using the DREAM.3D software package \cite{Groeber2014} and then took slices of these three dimensional (3D) representations.
For ease of slicing and subsequent simulation efficiency, the cubes were 32$^3$ element structured meshes with voxelated grain boundaries.
To minimize correlation between 2D samples we sliced each cube on every 7th plane in the third dimension.
For the 2D slices only the in-plane texture was preserved so that the crystal orientation can be described by the in-plane rotation by angle $\pth$ of a $\langle 100 \rangle$ canonically oriented FCC crystal.
This procedure has the benefit of being connected to, if not a full representation of, realistic 3D polycrystals.

\begin{figure}
\centering
\includegraphics[width=0.7\textwidth]{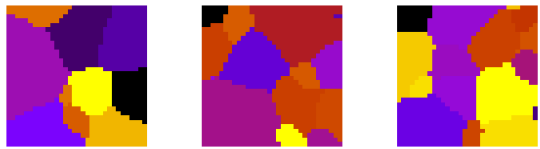}
\caption{Oligocrystalline realizations colored by crystal orientation.
}
\label{fig:realizations}
\end{figure}

\begin{figure}
\centering
\subfloat[angle distribution]
{\includegraphics[width=0.45\textwidth]{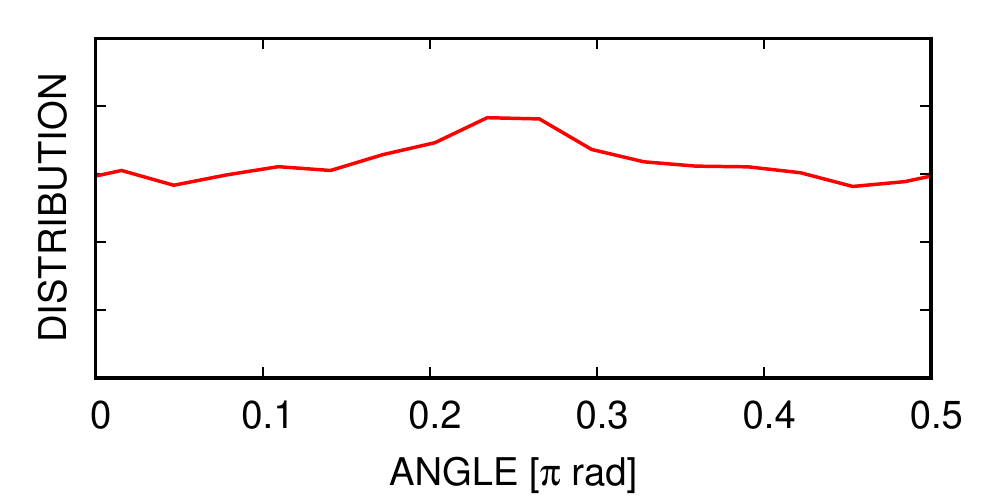}}
\subfloat[angle correlation]
{\includegraphics[width=0.45\textwidth]{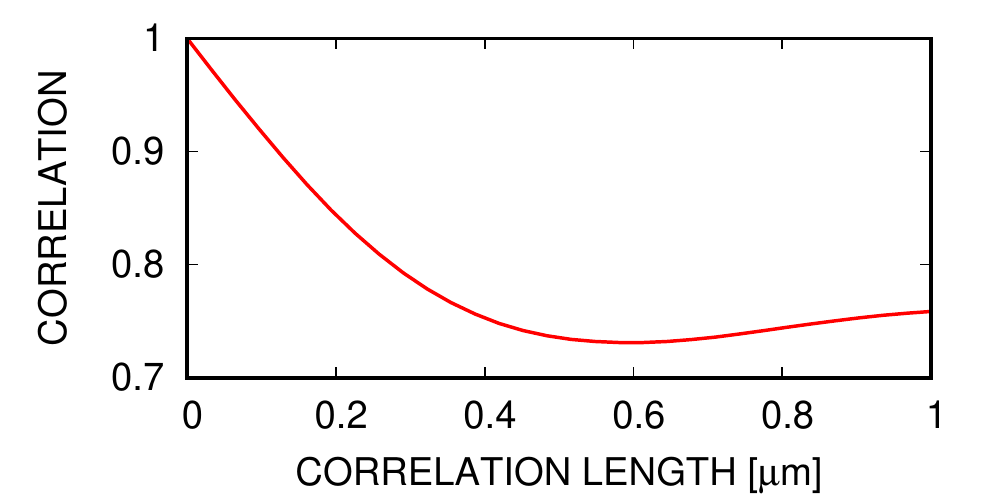}}

\subfloat[number of grains]
{\includegraphics[width=0.45\textwidth]{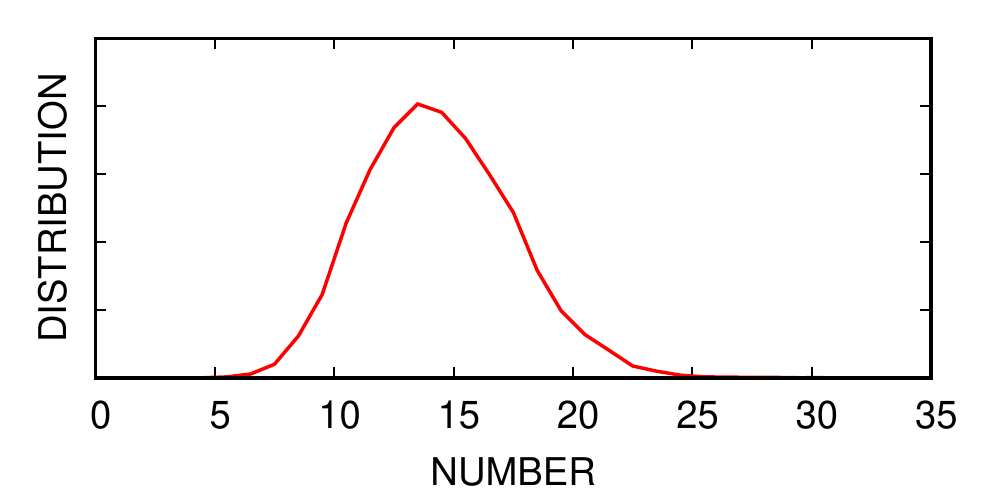}}
\subfloat[grain size]
{\includegraphics[width=0.45\textwidth]{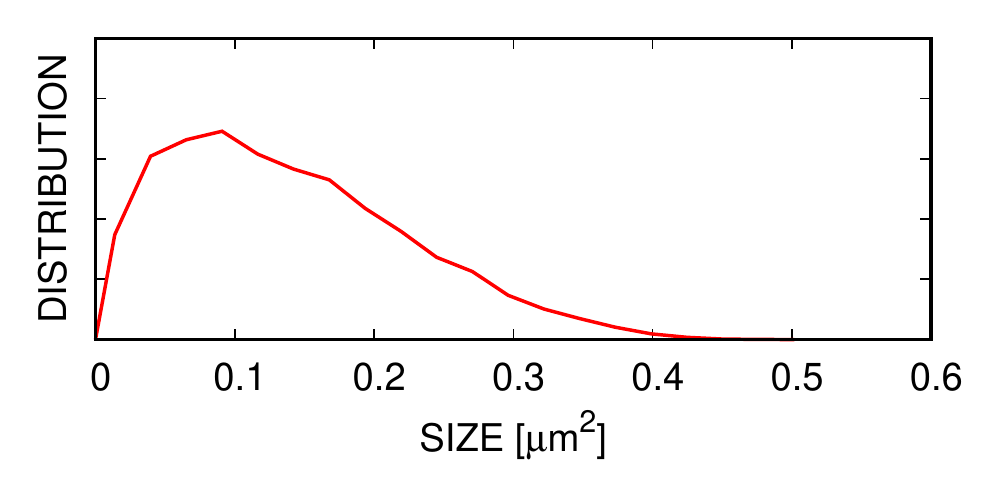}}

\subfloat[misorientation]
{\includegraphics[width=0.45\textwidth]{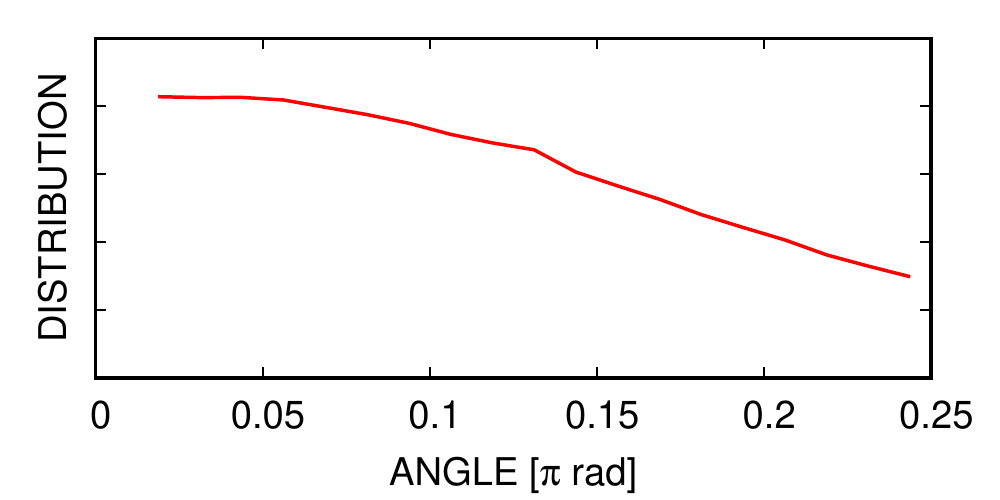}}
\subfloat[boundary length]
{\includegraphics[width=0.45\textwidth]{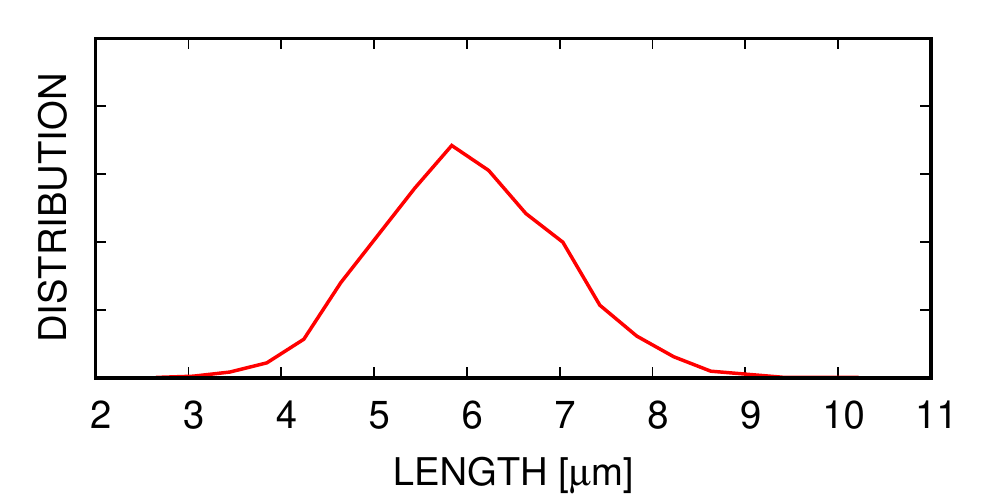}}
\caption{Descriptive statistics of the oligocrystal ensemble.
}
\label{fig:grain_stats}
\end{figure}

\fref{fig:realizations} shows representative realizations from this process colored by the texture angle $\phi \in [0,\pi/2)$ (black to yellow).
The pixelated grain boundaries are apparent, as are the fairly equiaxed grains.
\fref{fig:grain_stats} shows descriptive statistics of the ensemble of 2D realizations.
The orientation angle distribution is approximately uniform and is correlated over the spatial domain of the realizations.
Clearly the 2D  (and 3D) samples are stochastic volume elements (SVEs) since the correlations are long compared to the size of the domain.
This spatial correlation was in inherited from the process that created the 3D cubes.
Since the samples are not at the representative volume element (RVE) limit, we will refer to the realizations as oligocrystals (as opposed to polycrystals) in the subsequent sections.
The number of grains per realization ranges from 5 to 25, with 14 being the most prevalent (corresponding to a density of 0.071 $\mu$m$^{-2}$).
The distribution in grain size is approximately log normal with mode at approximately 0.1 $\mu$m$^2$.
Again, this distribution is inherited from the process used to create the 3D realizations.
The distribution of misorientation between grains has slight preference for low angle boundaries but there are also a significant number of high angle boundaries.
Lastly, the grain boundary length distribution is fairly compact with a mode at approximately 6 $\mu$m.

\subsection{Mechanical response}
Given the crystal orientation, the response of each grain in the oligocrystal followed an widely employed elastic-viscoplastic constitutive response \cite{taylor1934mechanism,kroner1961plastic,bishop1951xlvi,bishop1951cxxviii,mandel1965generalisation,dawson2000computational,roters2010overview} representative of austenitic (FCC) steel.
For the crystal elasticity, we employed a finite deformation Saint Venant model, which is linear relation between the second Piola-Kirchhoff stress $\Sb$ in the intermediate configuration and the elastic Lagrangian strain $\Eb^e$,
\begin{equation}
\Sb = \Cbb : \Eb^e \ .
\end{equation}
The Cauchy stress $\boldsymbol{\sigma}$ is obtained through the usual relation involving the deformation gradient.
Since elastic modulus tensor $\Cbb$ respects FCC crystal symmetries, there are only three relevant non-zero components, $C_{11}, C_{12}, C_{44}$, and these were set to 204.6, 137.7, 126.2 GPa, respectively.
In 3D, plastic flow can occur on any of the 12 FCC slip planes (with Schmidt tensors $\{\sb\otimes\nb\} = (111)\otimes\langle 110 \rangle$) in each crystal subject to compatibility between grains.
The allowed in-plane rotation left the third component of the slip $\sb$ and normal $\nb$ vectors unchanged; and, a plane strain constraint was accomplished via projection of the motion, \ie removal of the third component of the slip.
As in a previous study \cite{frankel2019predicting}, we employed a power-law form for the slip rate relation
\begin{equation}
\dot{\gamma}_{\alpha}=\dot{\gamma}_0\left|\frac{\tau_{\alpha}}{g_{\alpha}}\right|^{m-1}\tau_{\alpha} \ ,
\end{equation}
driven by the shear stress $\tau_\alpha$ resolved on slip system $\alpha$.
We chose the reference slip rate $\dot{\gamma}_0$ = 1.0 s$^{-1}$, the rate exponent $m = 20$, and the initial slip resistances $g_{\alpha}(t=0)$ = 122.0 MPa.
The slip resistance evolves according to \cite{Kocks1976, mecking1976hardening}
\begin{equation}
\dot{g}_\alpha = (H-R \, g_\alpha) \sum_\alpha |\dot{\gamma}_\alpha|
\end{equation}
where the hardening modulus is $H = 355.0$ MPa and the recovery constant is $R = 2.9$.
See \cref{jones2019machine,frankel2019predicting} for more details.

\begin{figure}
\centering
\includegraphics[width=0.4\textwidth]{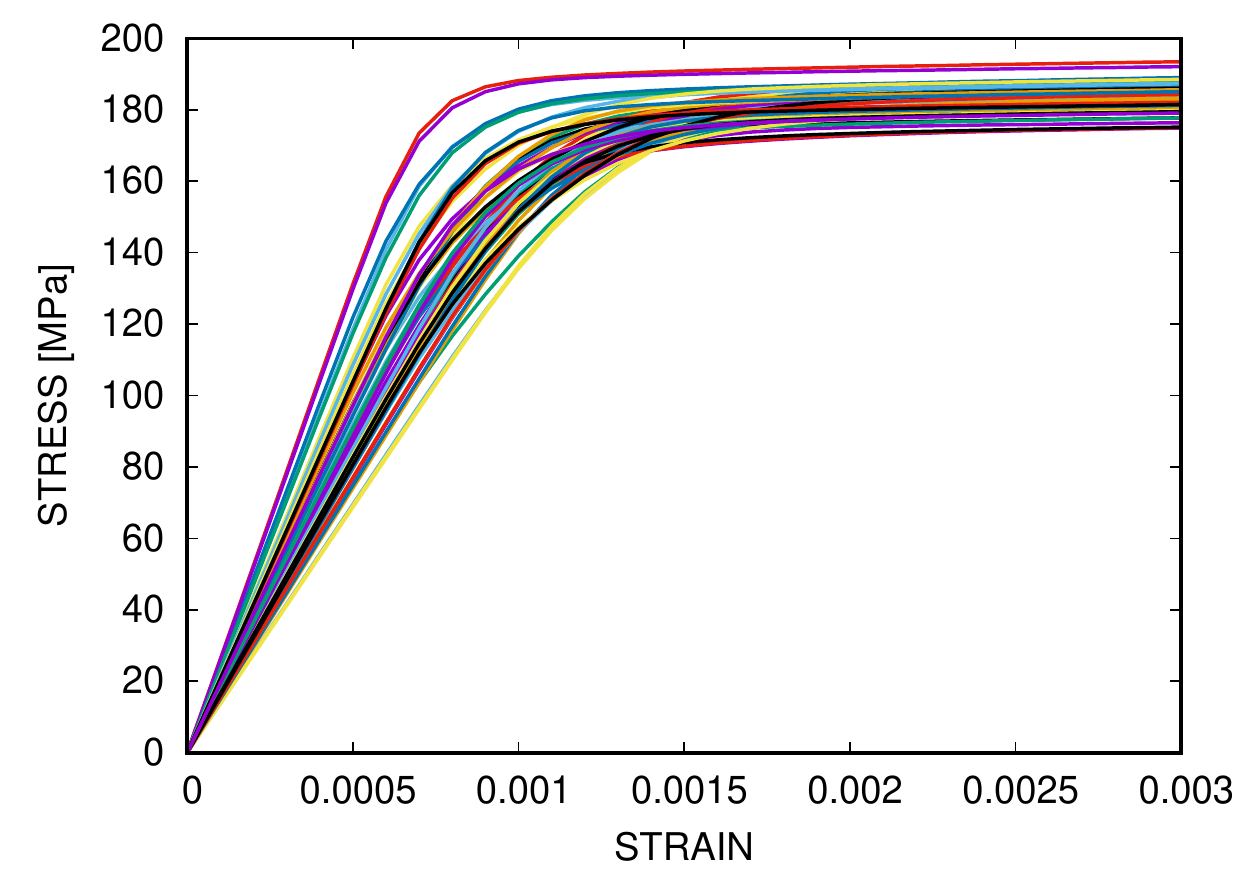}
\caption{Stress-strain response for a number of realizations.
}
\label{fig:stress_data}
\end{figure}

Under the plane strain constraint, each oligocrystal realization was subjected to quasi-static uniaxial tension at a constant engineering strain-rate of $\dot{\epsilon} = 1$ $s^{-1}$ up to 0.3\% strain with free lateral boundaries and uniform displacement on the top and bottom.
The loading range chosen to cover the elastic plastic transition and the subsequent plastic flow, as can be seen in the representative stress-strain data is shown in \fref{fig:stress_data} (representative stress fields are shown in \fref{fig:stress_field_evolution} and will be discussed in \sref{sec:results}).
These simulations relied on Albany \cite{salinger2016albany}.
Note that a richer set of loading conditions could be explored, at the cost of increasing the size and cost of the training data manifold \cite{jones2018machine}.
Here we focus on tension since this is the most common loading condition in experiments.

\section{Neural network model} \label{sec:nn}

Given the relatively recent application of machine learning to materials science, we first give a brief overview of the relevant techniques, specifically neural networks that can be applied to predicting the evolution of the stress field in an oligocrystal given an initial grain structure and the loading.
Then we describe the particular the architecture and training of the proposed network.

The basic type of artificial neural network (NN) is the {\it multilayer perceptron} (MLP) \cite{rosenblatt1961principles}.
This feed-forward model is a directed graph that feeds the selected inputs through layers of nodes that are densely connected between neighboring layers.
Each node transforms its inputs via a non-linear activation function with trainable weights and biases.
The result of all the transformations of the inputs is given by the output layer that is compared to data in training/calibrating the model using a selected error/loss metric.
A MLP neural network with $L$ layers and $N$ nodes per layer requires the optimization of $O(LN^2)$ parameters.

For image or field data, the number of parameters needed to model a function of the image is too large to be trained practically, and would require a prohibitive amount of data in the current materials science context.
However, for field data, it is physically plausible that spatial correlations exist that decay with distance.
The {\it convolutional neural network} (CNN) \cite{goodfellow2016deep} was developed to take advantage of this locality.
Rather than prescribing a unique set of weights to every single pixel, the convolutional neural network uses a convolutional filter with compact support and single set of weights that is moved across the entire image to product output of the same dimension.
The result of this operation is a new image and multiple such sets of filters can be applied in parallel.
In general, further non-linear operations, reductions of the image, or convolutions may be performed to enhance the richness of the feature space of the image or reduce the dimensionality before finally yielding the desired output.

A similar heuristic is commonly used for time-series data, where a long sequence data may require an inordinate number of parameters to fit to an output.
Based on causality, we expect that a future state will depend most strongly on the current state along with some additional latent variables controlling non-stationarity of the signal being modeled.
Thus {\it recurrent neural network} (RNN) architectures were created to take in as input the input features from the current state as well as the output from the previous state as inputs to a single layer MLP in order to make predictions.
In this sense, they have some features in common with traditional time integrators.
A number of advanced variants of the recurrent neural network design, including the {\it long-short-term memory} (LSTM) \cite{hochreiter1997long}, include additional hidden variables that take into account a longer running history of the signal and thus improve the accuracy and stability of training these neural networks.

The {\it convolutional LSTM} (ConvLSTM) \cite{xingjian2015convolutional} was designed to combine the heuristics of the CNN and LSTM architectures for making predictions of a sequence of images.
Where an LSTM uses a small MLP as a non-linearity for feature processing and prediction, a convolutional LSTM uses a set of convolution operations on each image to create the next image; the input image from the current state, as well as the output image from the previous state and hidden variables, are used as inputs to make subsequent predictions.
This architecture has particular appeal for microstructural mechanics since it exploits both spatial and temporal correlations in the data to be modelled.
For our application, the physical topology of the grains influences the stress state of the microstructure in a way that simple reduced order statistics like average grain size cannot capture reliably, and so we need a model that can analyze the full field physical state of the system at once.
At the same time, the evolution of the stress as a function of strain is expected to depend heavily on the previous stress states as well as the evolution of the latent variables associated with plastic dissipation, which indicates a sequential model with a longer-term memory.

\subsection{Architecture}

\begin{figure}
\centering
\includegraphics[width=0.55\textwidth]{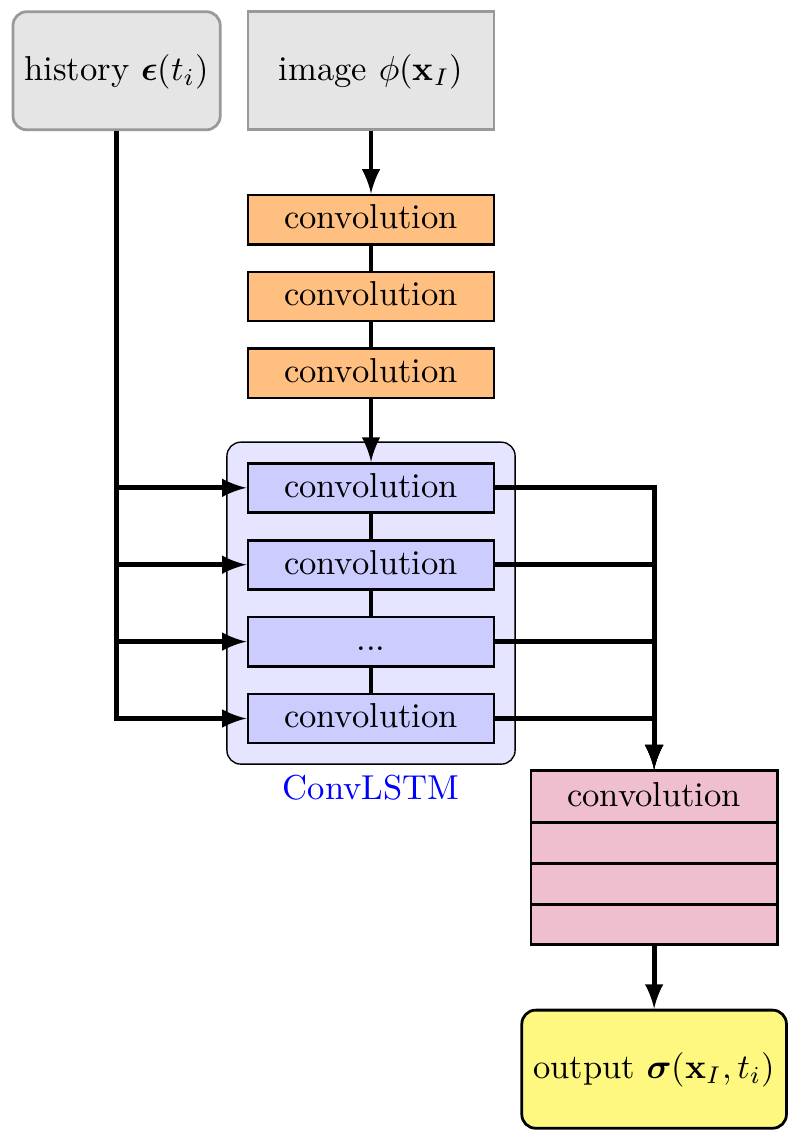}
\caption{Neural network architecture.
Input is (a) the external strain loading of the oligocrystal $\strain(t_i)$ and (b) an image of the initial microstructure (orientation field) $\pth(\xb_I)$.
The image $\pth(\xb_I)$ is fed to a sequence of convolutions (orange layers).
This output of all the filtered images associated the last convolution is given to a sequence of recurrent layers in the ConvLSTM (blue) together with loading.
At the end, a block 3D convolution (red) takes the image sequence from the ConvLSTM to generate the output stress evolution $\stress(\xb_I,t_i)$ (yellow).
}
\label{fig:nn_architecture}
\end{figure}

\fref{fig:nn_architecture} shows a schematic of the proposed architecture which has a ConvLSTM at its core.
The inputs (gray) to the network are: (a) a time sequence $\strain(t_i)$, and (b) an image $\phi(\xb_I)$ (2D array) corresponding to the first time, $t_0$.
The outputs (yellow) are images, $\stress(\xb_I,t_i)$, for each time $t_i$ in the sequence (a 3D array).
Unlike in traditional applications where the input image and output images are of the same quantity, in our network the initial image $\phi(\xb_I)$ is the crystal orientation field characterizing texture, while the output images are the stress field $\stress(\xb_I,t_i)$ as a function of time/external loading strain $\strain(t_i)$.
Here, $\xb_I$ are element locations which are identified with image pixels $I$ and $\phi(\xb_I)$ with the color of the pixel $I$.
Image size $N\times N$ = 32$\times$32 is fixed by the data described in \sref{sec:data}.
The image of the initial microstructure is fed into a sequence of $N_\text{layers}$ 2D convolutional layers (Conv2D, orange) with rectified linear unit (ReLU) non-linear activations \cite{resteghini2013single}.
Each layer has $N_\text{filters}$ filters applied in parallel.
Each of the filters has a kernel size of 4$\times$4, with zero-padding to maintain the image size pre-to-post filter.

The output of the initial convolutional layers and the loading history is combined in the ConvLSTM (blue); its recursive structure is shown unrolled in \fref{fig:nn_architecture} for illustration.
To embed the time-dependence of the problem, each processed image is repeated for each of the time steps, and an additional filter is added to each whose value is constant across the image and equal to the value of the time step.
Thus the input to the convolutional LSTM is a 4-dimensional tensor of size $N_\text{times}\times N \times N \times (N_\text{filters}+1)$, indexed by time step, image row, image column, and filter value, respectively.
The ConvLSTM used the same kernel size, number of filters, padding, and activation as the initial 2D convolutional layers.
Its output was sequence of images, one for each time step.
Finally, a 3D convolutional layer (Conv3D, red) with 1 filter was applied across the two space and one time dimensions of the input data, as processed by the ConvLSTM.
Its output (yellow) is the time sequence of images of the stress field $\stress(\xb_I,t_i)$.
To keep the cost of training manageable, the stress field data was limited to the normal component of stress in the tension direction \ie the most significant component.

The Keras-Tensorflow framework \cite{tensorflow} was used to construct and train the proposed neural network architecture.

\subsection{Training} \label{sec:training}

To train the proposed network and some variants, a dataset of 16,000 input-output pairs $\{ [ \strain(t_i, \phi(\xb_I) ], \stress(\xb_I,t_i) \}$ from the CP simulations described in \sref{sec:data} was split into 80\% for training and 20\% for testing.
One tenth of the training set was used for network validation to monitor for convergence during training and prevent over-fitting.
This amounts to 11,520 training samples, 1,280 validation samples, and 3,200 testing samples.
The networks were trained using an Adam optimizer \cite{kingma2014adam} guided by a mean-squared-error loss function summed over all pixels $I$ and times $t_i$.
The Adam optimizer \cite{kingma2014adam} used an initial learning rate of 0.001 and a batch-size of 128.
GPUs were used to overcome the computational expense of training the network so that each training epoch took approximately 100 seconds.
Training proceeded for 1500 epochs or until the validation error stopped decreasing.
To aid in the training, the stress field data was also normalized such that the average value of the stress at the final time step was equal to 1.

\begin{figure}
\centering
{\includegraphics[width=0.55\textwidth]{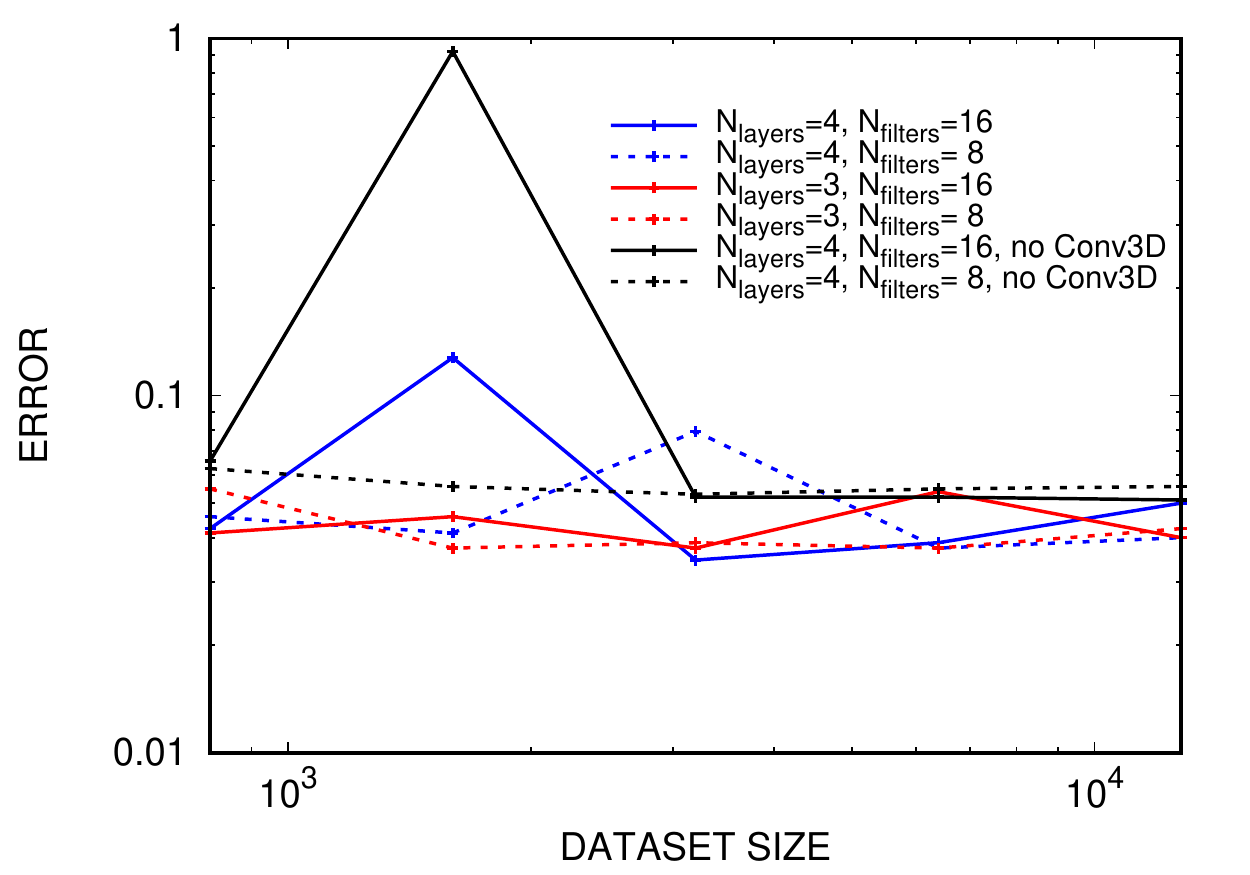}}
\caption{Convergence with dataset size
(a) baseline ($N_\text{filters}$=16,  $N_\text{layers}$ =4)
(a) baseline with fewer filters ($N_\text{filters}$=8),
(b) baseline with fewer input convolutional layers ($N_\text{layers}$ =3),
(c) baseline without the Conv3D layer so that the ConvLSTM directly outputs the stress evolution predictions.
}
\label{fig:convergence}
\end{figure}

The network shown in \fref{fig:nn_architecture} has large space of hyperparameters, especially if we were to assign kernel width, \etc per layer.
We did a non-exhaustive exploration of the hyperparameter space to determine the relative performance of variants of the architecture.
\fref{fig:convergence} compares the root mean square (pixel-wise) validation error versus size of training dataset for a few architecture variants.
For each of the trainings the training and testing errors were comparable, which indicates that the networks were not overfit to the data.
The flattening of the error curves in \fref{fig:convergence} is evidence of limits to the learning, which indicates that the networks were not complex enough or the increase in the size of the dataset provides no additional information.
The main finding of the architecture exploration is that the Conv3D appears to be needed to reduce the training error.
For the other modifications, namely changing the number of Conv2D layers and the number of filters, the results are more equivocal.
Since we were only able to train each network variant once due to the training costs, we conjecture that the scatter in the losses of the best network is likely due to the vagaries of the stochastic optimization method (and the early stopping criterion).
Hence, we chose the architecture with $N_\text{filters}=16$ and $N_\text{layers}=4$ since it achieved the lowest error over all the dataset sizes.
Further optimization of the network architecture should be possible.

\section{Results} \label{sec:results}

After training the model we evaluated the quality of its predictions and demonstrated its use in material evaluation/optimization.

\subsection{Quality of the predictions}

We using the trained model to predict the response of the 2,400 samples remaining from train/test split (described in \sref{sec:training}) and compared the stress history $\stress(\xb_I,t)$ predictions to the corresponding held out CP data.
Note that, given the strain rate, the time $t$ and the loading strain $\strain(t)$ have the same values.

\begin{figure}
\centering
\includegraphics[width=0.45\textwidth]{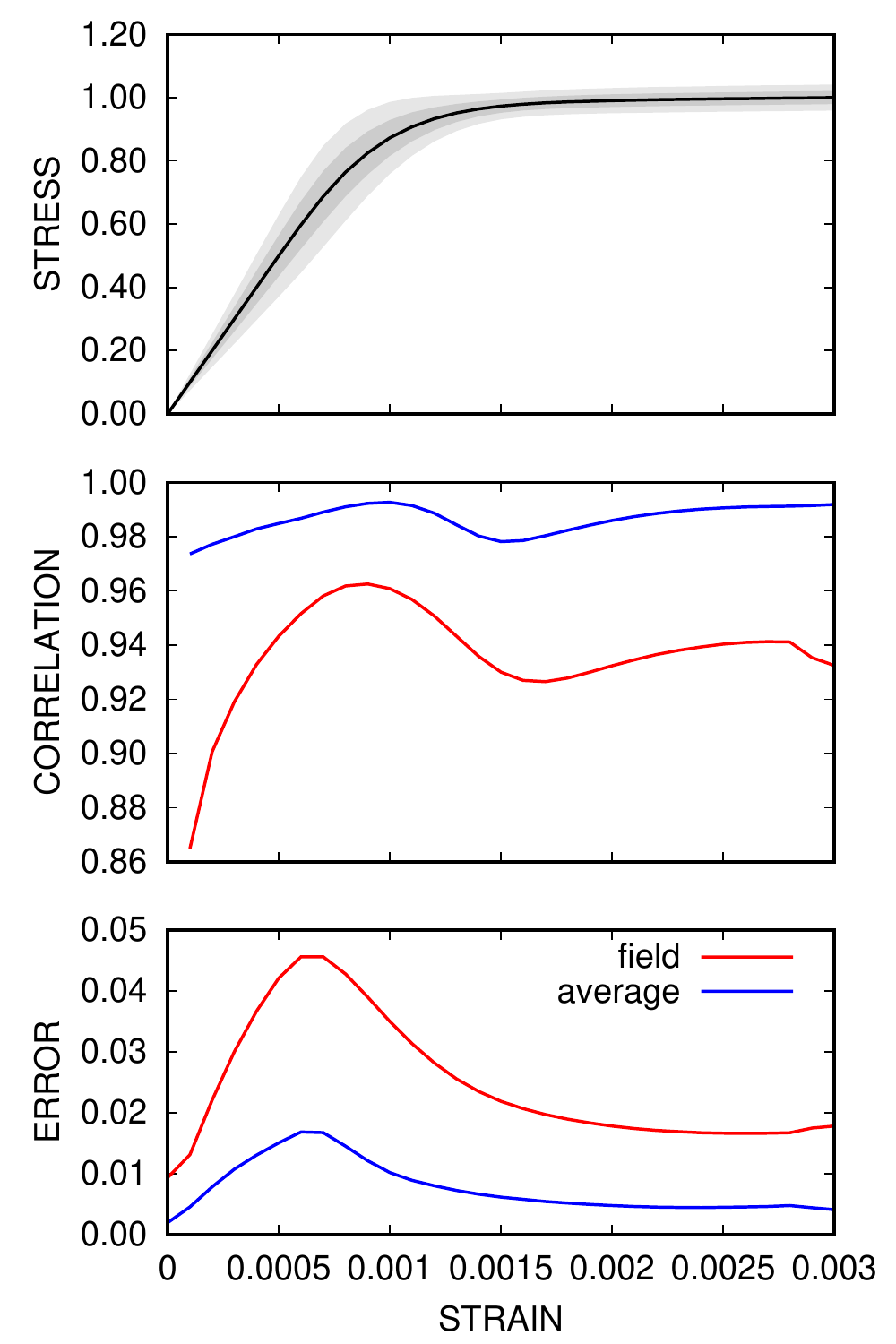}
\caption{Normalized stress response (upper panel), correlations (middle panel), and root mean squared errors (lower panel) versus strain.
Upper panel shows mean response (black line), and one (dark gray) and two (light gray) standard deviations with strain
Middle panel correlation of predictions for full field (red) and average (blue).
Middle panel mean squared error of predictions for full field (red) and average (blue).
}
\label{fig:correlation_error}
\end{figure}

\fref{fig:correlation_error} shows the correlation of the NN predictions with the true/CP stresses and the root mean squared (pixel-wise) error of the normalized stresses.
The stresses are normalized by the maximum of their over the strain history.
Both the correlation of the spatial average of $\stress(\xb_I,t)$ and the full field correlations and errors are shown.
The spatial average displays a cancellation of errors and better correlation with data than the full field but both are good.
The ensemble-average stress-strain response with ensemble variance is shown in the top panel to aid comparison with trends in the correlations and errors.
The absolute error in the normalized stress for both the average and the field are worse slightly preceding the elastic-plastic transition, where there is the highest variance in the ensemble response and the most complexity in the local states.
Interestingly the correlation coefficient was highest around the elastic-plastic transition, which implies model bias in this regime.
Unlike our previous work \cite{frankel2019predicting} which displayed monotonic increase in error with time with a MLP-based LSTM architecture used to predict the spatial average stress history, the Conv3D structure in the present NN appears to level and reduce the error over the entire history (as well as effectively predicting the evolution of the entire field).
However, some artifacts of the zero padding in time can be seen in the error curves in \fref{fig:correlation_error} and the predicted (zero load) first stress states are not identically zero (not shown).

\begin{figure}
\centering
\subfloat[]
{\includegraphics[width=0.90\textwidth]{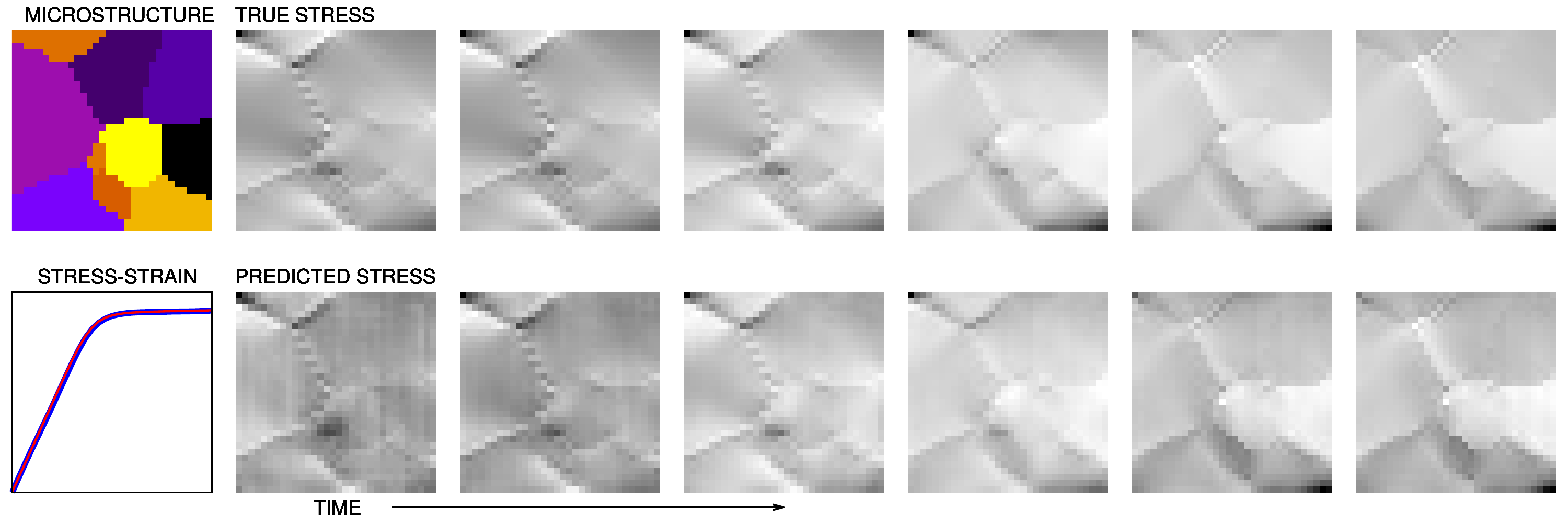}}

\subfloat[]
{\includegraphics[width=0.90\textwidth]{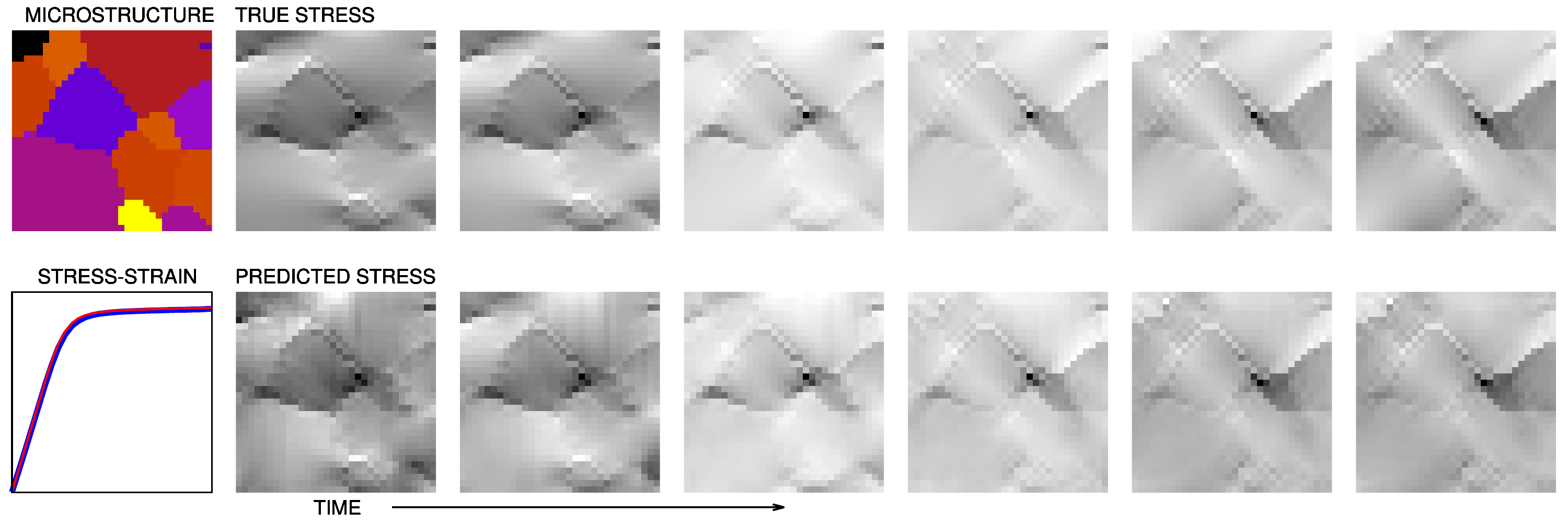}}

\subfloat[]
{\includegraphics[width=0.90\textwidth]{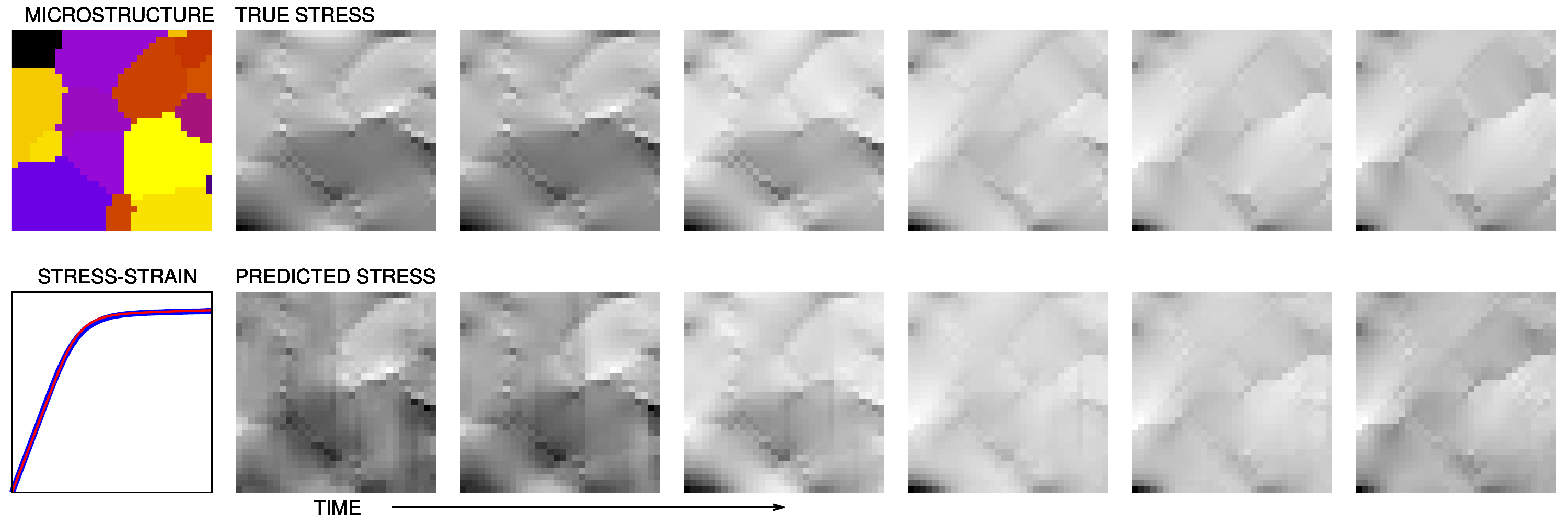}}
\caption{Evolution of stress.
Three representative realizations (a,b,c):
initial grain structure (color,left upper panel),
comparison of the oligocrystal average stress-strain response (CP:blue, NN:red, left lower panel), and
stress field evolution (gray, true/CP:upper right panels,predicted/NN:lower right panels) at uniformly sampled times $t$ across the range $t \in [0,0.003]$.
}
\label{fig:stress_field_evolution}
\end{figure}

\begin{figure}
\centering
\subfloat[]
{\includegraphics[width=0.90\textwidth]{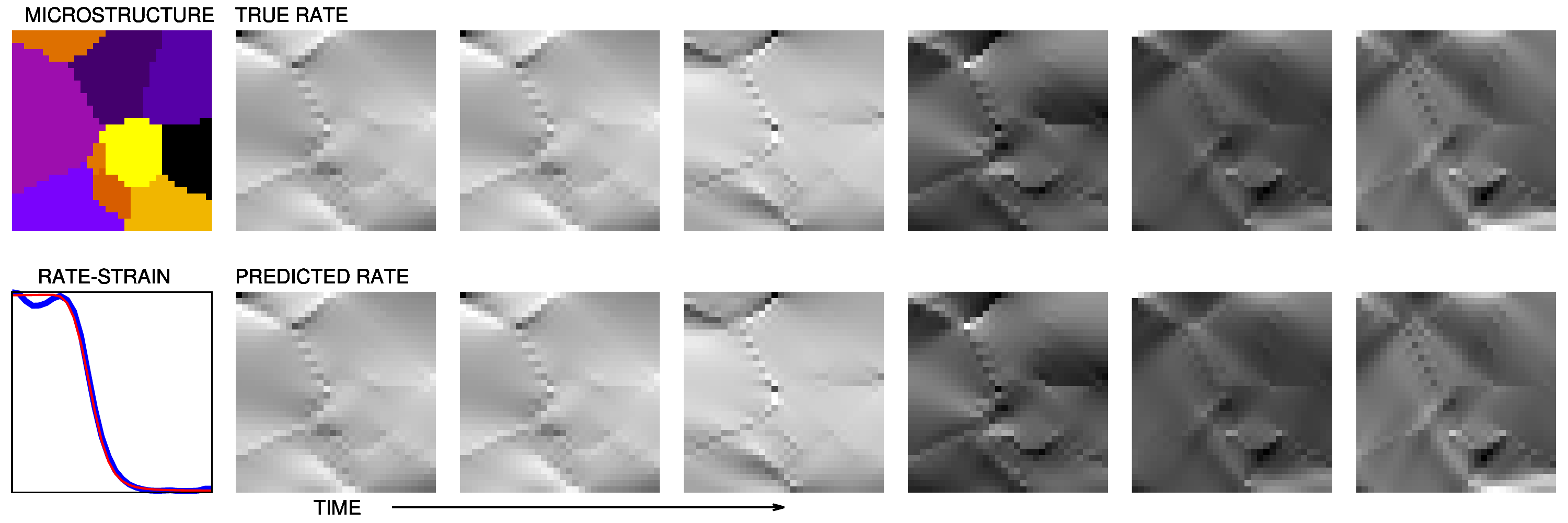}}

\subfloat[]
{\includegraphics[width=0.90\textwidth]{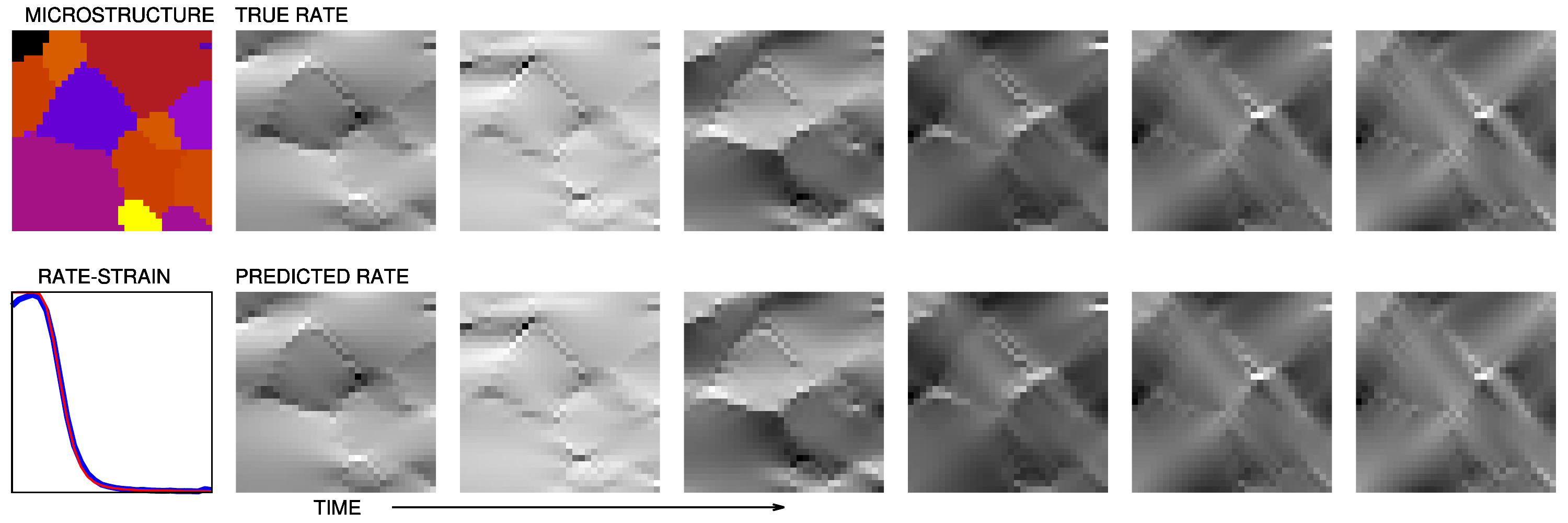}}

\subfloat[]
{\includegraphics[width=0.90\textwidth]{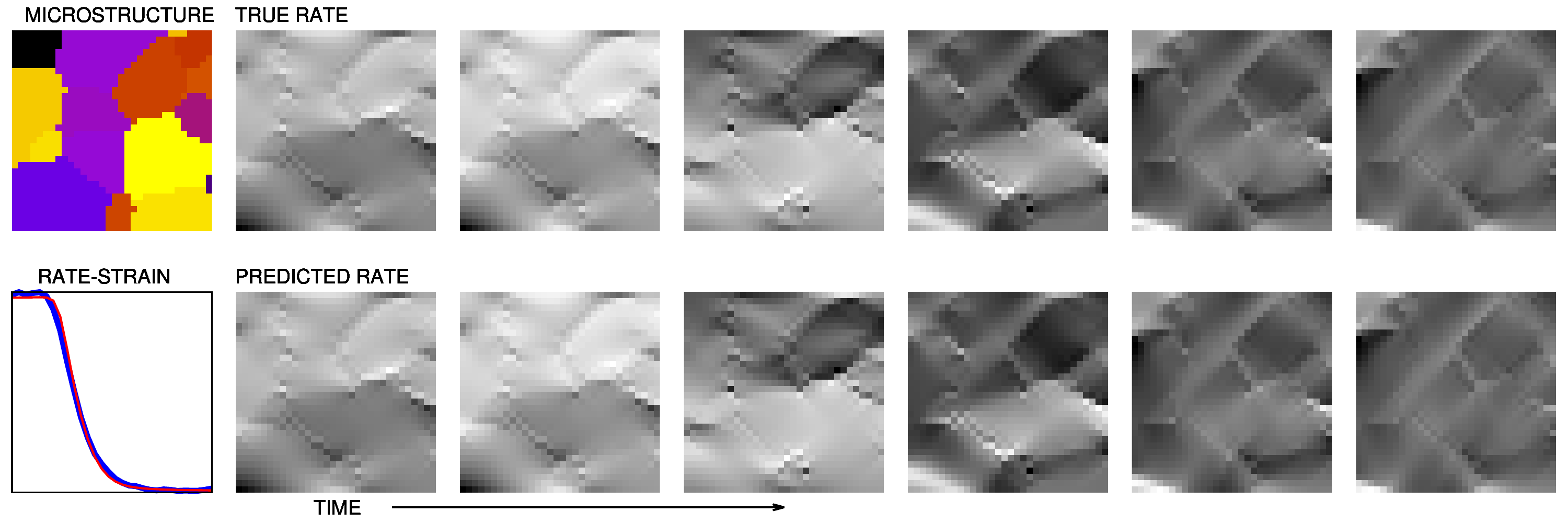}}
\caption{Evolution of stress rate.
Three representative realizations (a,b,c):
initial grain structure (color,left upper panel),
comparison of the oligocrystal average stress-strain response (CP:blue, NN:red, left lower panel), and
stress field evolution (gray, true/CP:upper right panels,predicted/NN:lower right panels) at uniformly sampled times $t$ across the range $t \in [0,0.003]$.
}
\label{fig:stress_rate_evolution}
\end{figure}

\fref{fig:stress_field_evolution} demonstrates that the predictions of the normal stress field evolution are high fidelity, with very few visible artifacts in the sequence of images (upper rows: true/CP data, lower rows: predicted/NN).
The correspondence is corroborated by the numerical correlation shown in \fref{fig:correlation_error}.
As mentioned, the initial prediction at $t=0$ (not shown for brevity), before any deformation has occurred, is not identically zero; this could be corrected with an alternative structure for the NN or padding with more appropriate values, or minimized with particular weighting or penalty in the loss function.
These errors, as the errors throughout the history are arguably negligible since the mean response (predicted/NN (blue), true/CP (red)) lie on top of each other.
Both the predicted and true stress fields (grayscale image sequences) show stress concentrations in correspondence with the boundaries of the grain structure (color).
Predicting the stress rates is arguably a more difficult task since they all tend to zero as the plastic flow ensues.
\fref{fig:stress_rate_evolution} shows that the network predictions of the full field stress rates are in good correlation with the CP data and, given their positivity, satisfy the basic dissipation requirement implied by the second law of thermodynanmics.
Nevertheless, there are noticeable errors in the average rates early on in first and second realizations shown but not the third.

\begin{figure}
\centering
{\includegraphics[width=0.45\textwidth]{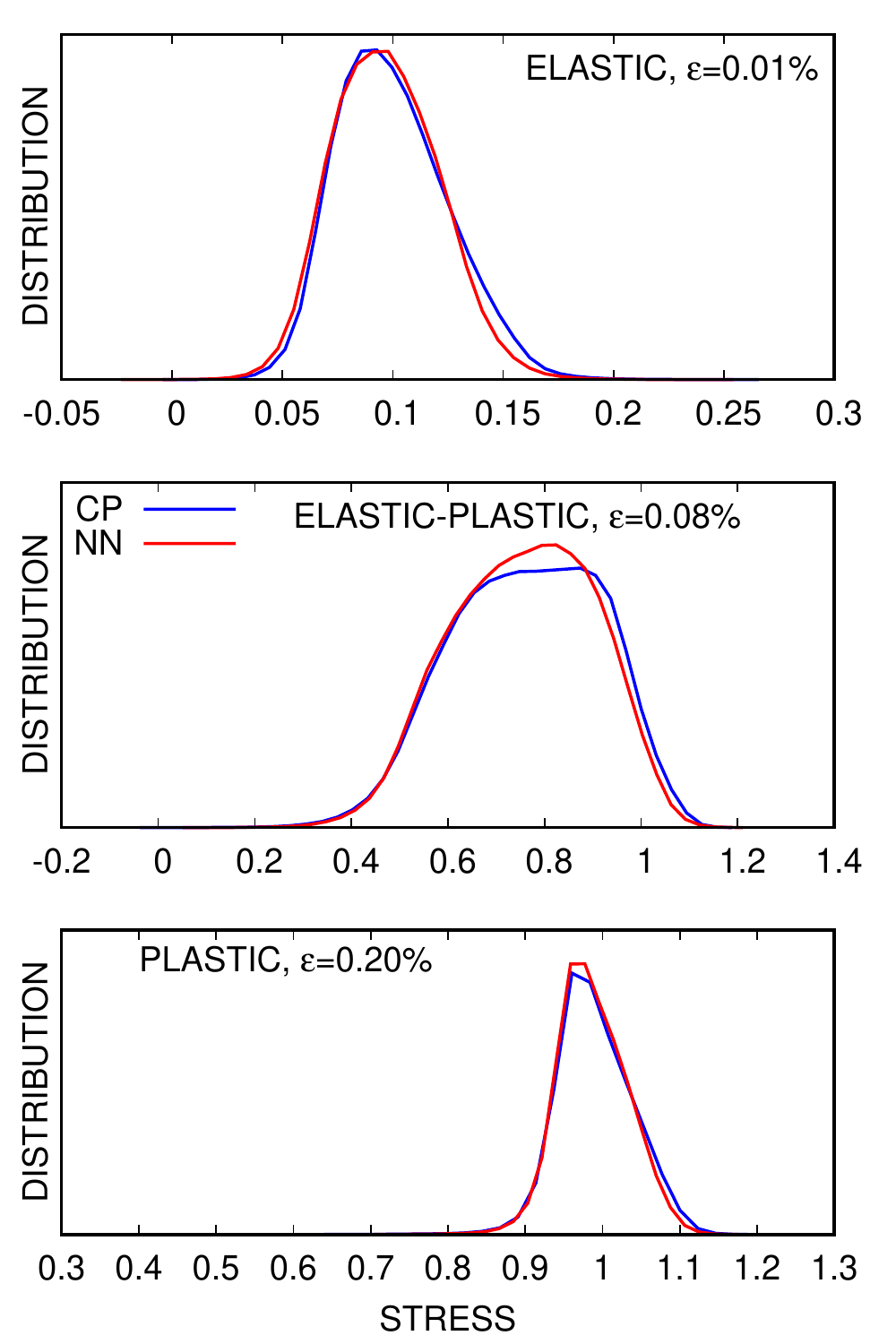}}
\caption{Evolution of the distribution of stress:
elastic regime (upper panel),
elastic-plastic transition (middle panel), and
plastic flow regime (lower panel).
}
\label{fig:stress_distribution_evolution}
\end{figure}

Lastly, we also compared the distributions of per pixel stresses over the ensemble at fixed loading levels.
\fref{fig:stress_distribution_evolution} shows the stress distributions in the elastic, elastic-plastic and plastic regimes.
Clearly, they are nearly identical.
The discrepancy for the elastic-plastic regime is the highest, as expected, since this is the most complex regime of the simulations.

\subsection{Structure-property exploration}

\begin{figure}
\centering
\subfloat[yield distribution]
{\includegraphics[width=0.45\textwidth]{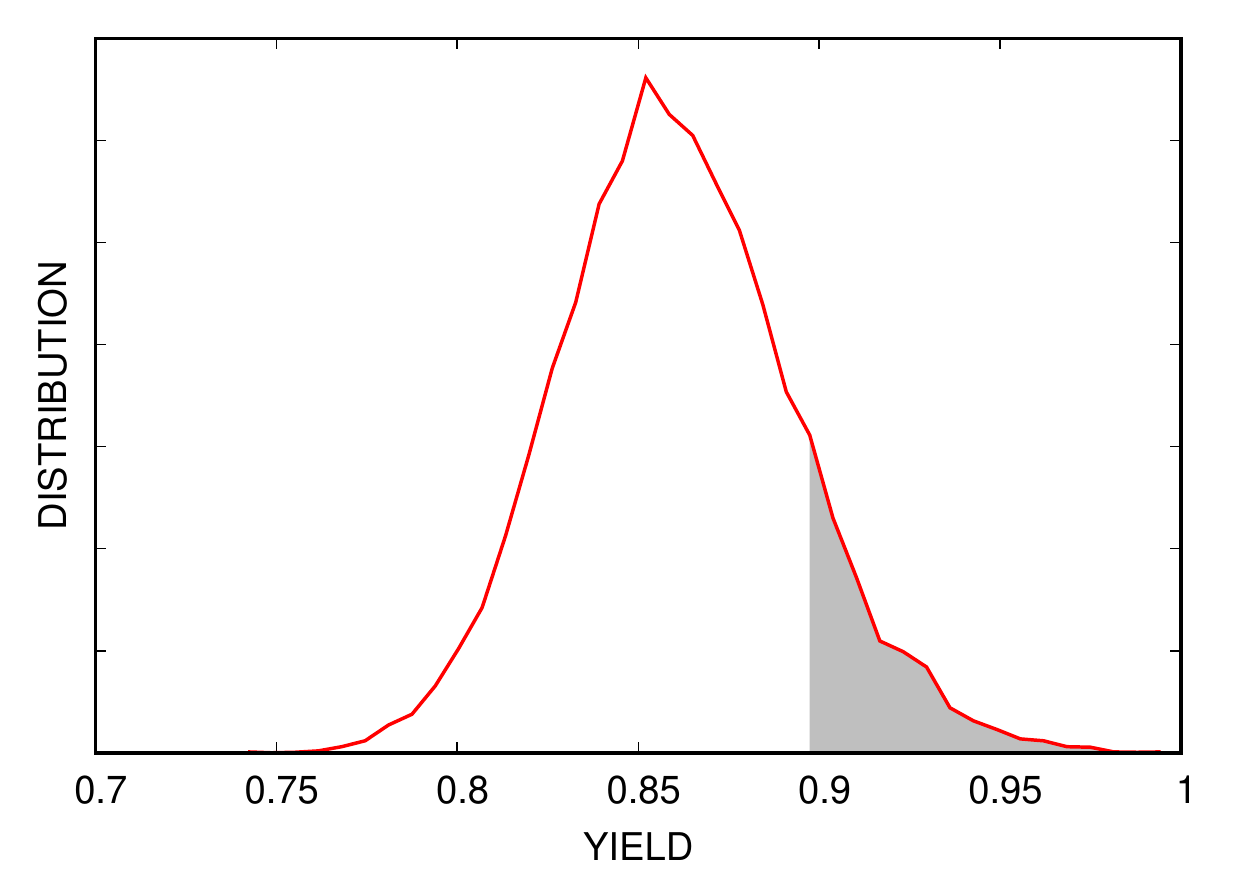}}
\subfloat[stress variance distribution]
{\includegraphics[width=0.45\textwidth]{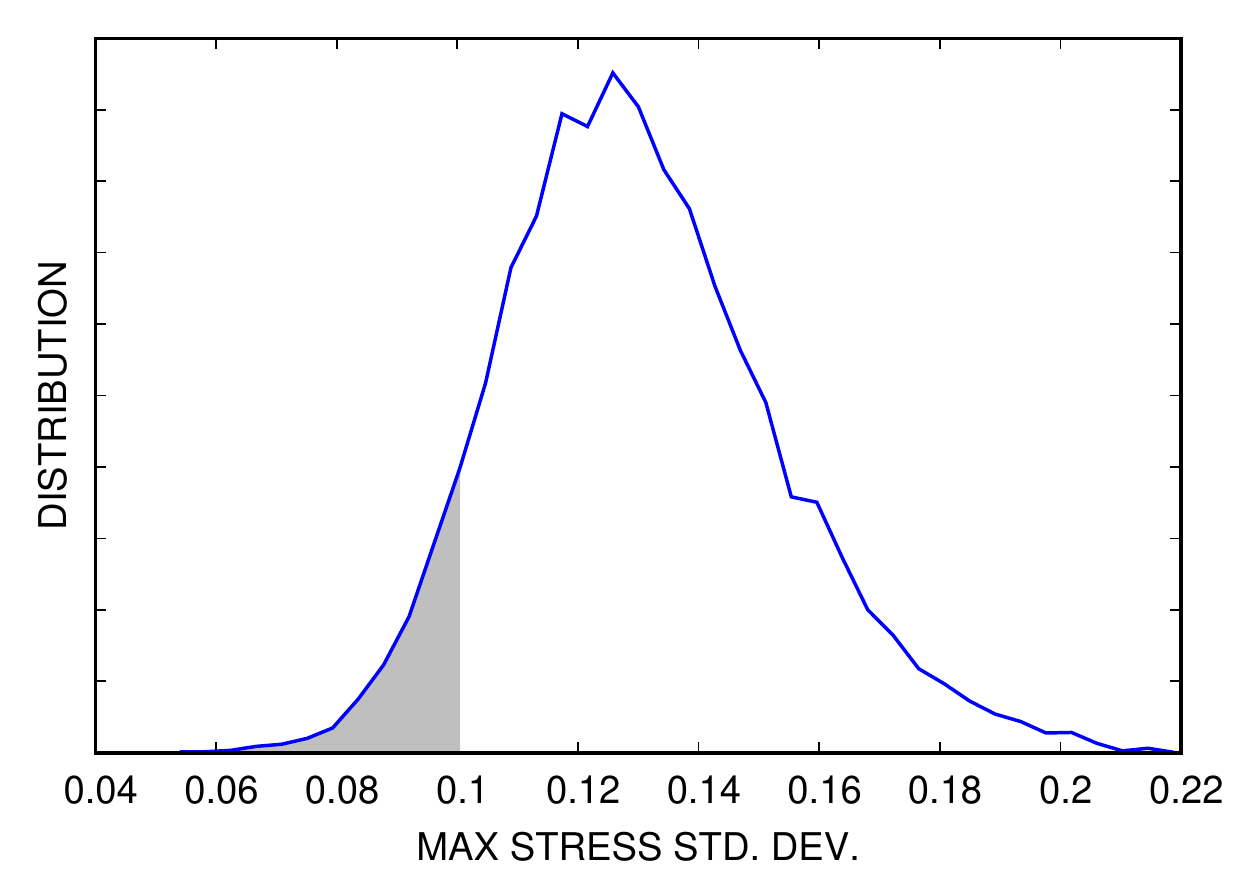}}

\subfloat[grain orientation distribution]
{\includegraphics[width=0.45\textwidth]{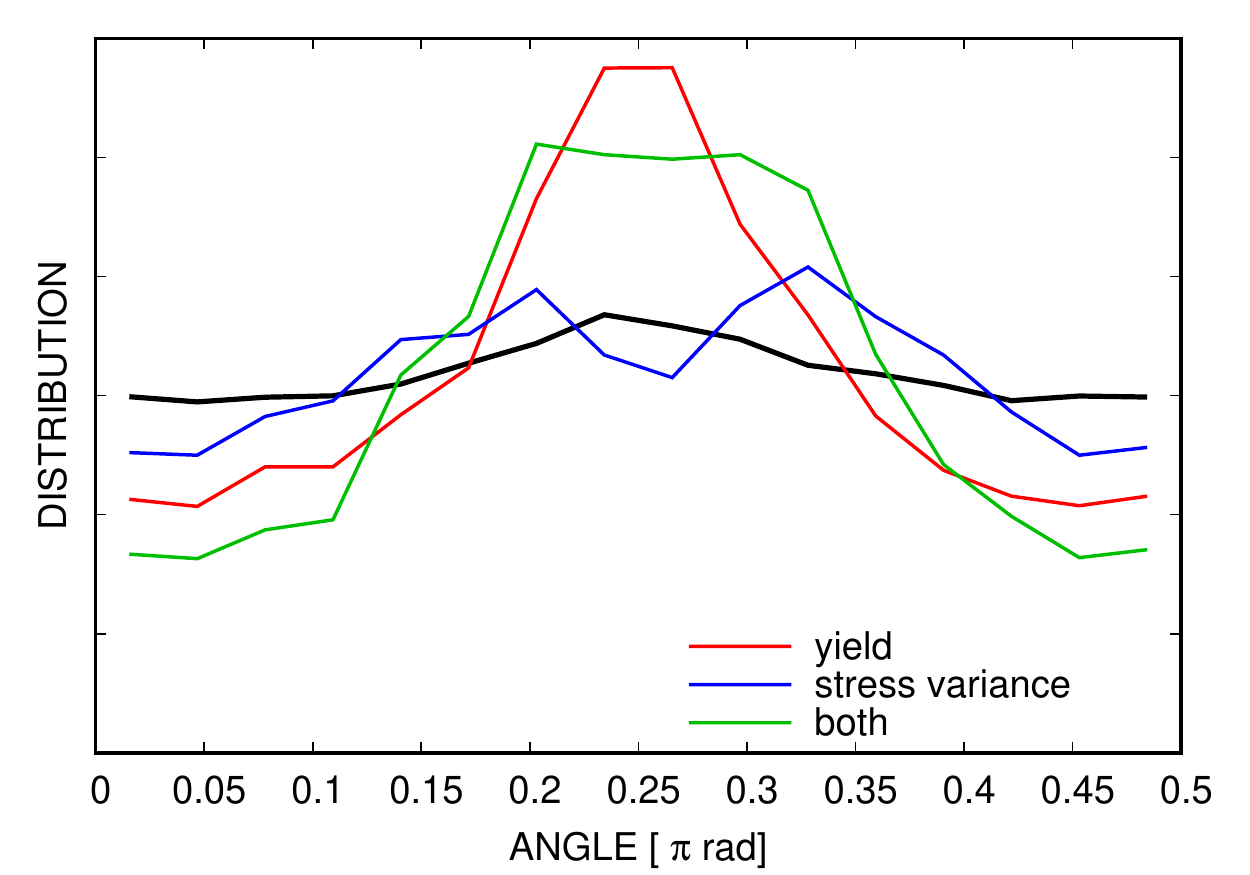}}
\subfloat[grain boundary length distribution]
{\includegraphics[width=0.45\textwidth]{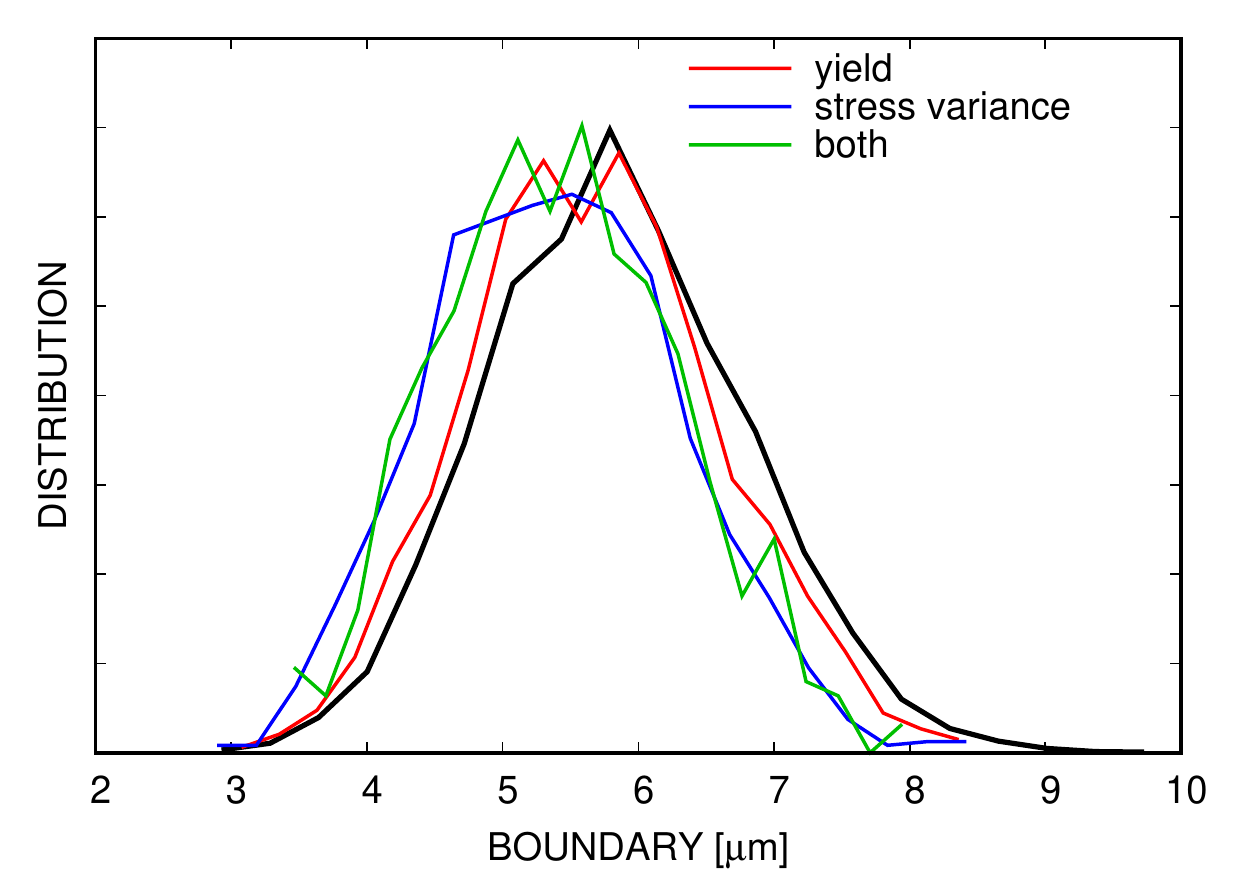}}
\caption{Screening of property distributions:
(a) yield strength distribution based on the normalized stress (gray: high yield population),
(b) distribution of maximum stress variation (gray: low variance population),
(c) distribution of grain orientation angles (black: full population/all realizations), and
(d) distribution of grain boundary length (black: full population).
}
\label{fig:opt_distribution}
\end{figure}

Using 16,000 additional oligocrystals generated with the same process as for the calibration, we used the NN model to screen these microstructures (\sref{sec:data}) based on their yield strength and the homogeneity of their stress fields.
Using thresholds shown in \fref{fig:opt_distribution}a,b for the yield strength and the maximum (over time) of the standard deviation the stress field, we were able to partition the ensemble into high yield structures (11.1\%), low variance structures (7.6\%) and structures that have both properties (2.4\%) in a few minutes.
To determine mechanistic sources for these properties we compared the distributions of the subpopulations to that of the full population for a variety of grain statistics.
\fref{fig:opt_distribution}c show interesting differences in the crystal textures in the various populations.
The high yield structures have a distinct prevalence of grains with $\phi=\pi/4$ orientations such that the slip planes are oblique to the loading direction.
The low variance structures have a bimodal distribution of textures with preference for near but off $\phi = \pi/4$ orientations.
The distribution of structures with both properties is unimodal but with broader peak than the high yield population.
\fref{fig:opt_distribution}d shows slight shift to lower grain boundary densities for the superior microstructures.
Since the misorientation was calculated such that it scales with boundary length, the distributions for the high yield and low variance populations also shifted to slightly lower values compared to distribution for the full population.
We also examined the distributions of grain density and grain size for the high performing structures.
They were were essentially equivalent with the distributions for the full population.

\begin{figure}
\centering
\subfloat[highest yield]
{\includegraphics[width=0.35\textwidth]{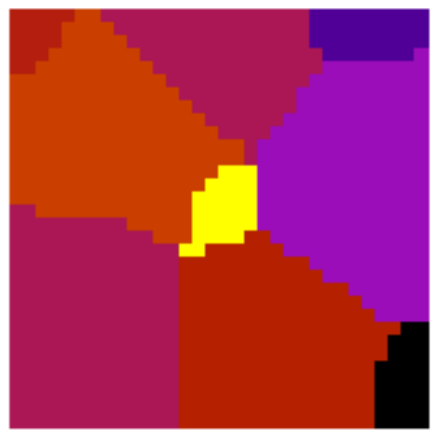}}
\subfloat[lowest stress variance]
{\includegraphics[width=0.35\textwidth]{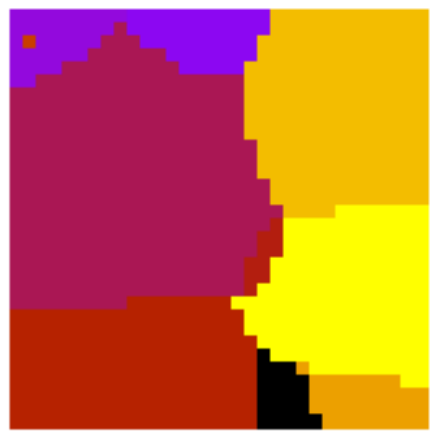}}
\caption{Best microstructures
(a) highest yield 0.998 in the low variance set,
(b) lowest stress variance 0.052 in the high yield set.
}
\label{fig:best}
\end{figure}

\fref{fig:best}a shows the highest yield microstructure and \fref{fig:best}b the lowest stress variance microstructure.
Although no properties of the structures are obviously apparent, \fref{fig:best}a seems to have a rough central symmetry and \fref{fig:best}b seems to have loose bilateral symmetry.
It is likely that these results are strongly influenced by boundary effects given the correlation length (refer to \fref{fig:grain_stats}b versus the size of the samples.

\section{Conclusion} \label{sec:conclusion}

Using a neural network architecture based on a ConvLSTM that efficiently encoded the spatial and temporal correlations in the data, we were able to predict the evolution the stress field of oligocrystals with high fidelity using just the initial microstructure and the external loading as inputs.
We also demonstrated that the network model can be used to facilitate material screening, optimization and design tasks.
A more sophisticated method would use the generator component of a GAN trained to generate microstructures \cite{yang2018microstructural} for material optimization over latent space/manifold embedded in the GAN.
In addition to structure-property discovery, the model can also be used as high efficiency surrogate model for uncertainty quantification and global sensitivity studies.
It is worth noting that computing the crystal plasticity training data took approximately 500 processor-days on a cluster, but the NN training took 2 days on a GPU, after which tens of thousands of evaluations of the NN model took minutes on a laptop.

Application of the proposed NN to other loading modes/non-monotonic loading should be straight forward, albeit with the significant sampling burden quantified in \cref{jones2019machine}.
In future work we will investigate the interpretability of the proposed network and would like to extend it to use multi-fidelity training augmented with experimental images and mechanical response.
Another topic of interest is the sensitivity of these types of network to the resolution of the grain structures and their spatial correlation.

\section*{Acknowledgments}
This work was supported by the LDRD program at Sandia National Laboratories, and its support is gratefully acknowledged.
Sandia National Laboratories is a multimission laboratory managed and operated by National Technology and Engineering Solutions of Sandia, LLC., a wholly owned subsidiary of Honeywell International, Inc., for the U.S. Department of Energy's National Nuclear Security Administration under contract DE-NA0003525.
The views expressed in the article do not necessarily represent the views of the U.S. Department of Energy or the United States Government.



\end{document}